\documentclass[10pt]{iopart}

\usepackage{bbm,iopams,mathrsfs,setstack}

\def\textbf#1{{\bf #1}}
\def\be{\begin{equation}}
\def\ee{\end{equation}}
\def\ben{\begin{eqnarray}}
\def\een{\end{eqnarray}}
\def\eea{\end{array}}
\def\bea{\begin{array}}
\newcommand{\ot}[0]{\otimes}
\newcommand{\bei}{\begin{itemize}}
\newcommand{\eei}{\end{itemize}}
\newcommand{\ket}[1]{|#1\rangle}
\newcommand{\bra}[1]{\langle#1|}

\newcommand{\proj}[1]{\ket{#1}\!\bra{#1}}

\def\blacksquare{\vrule height 4pt width 3pt depth2pt}

\newcommand{\supp}{\mathrm{supp}}
\newcommand{\spa}{\mathrm{span}}

\begin{document}

\title[A note on the optimality]{A note on the optimality of decomposable entanglement witnesses and completely entangled subspaces}

\author{R Augusiak$^{1}$, J Tura$^{2}$, and M Lewenstein$^{1,3,4}$}
\address{$^{1}$ICFO--Institut de Ci\`encies Fot\`oniques,
Parc Mediterrani de la Tecnologia, 08860 Castelldefels, Spain}
\address{$^{2}$Centre de Formaci\'o Interdisciplin\`aria Superior,
Universitat Polit\`ecnica de Catalunya, Pau Gargallo 5, 08028
Barcelona, Spain}
\address{$^{3}$ICREA--Instituci\'o Catalana de Recerca i Estudis Avan\c{c}ats, Lluis Companys 23, 08010 Barcelona, Spain}
\address{$^{4}$Kavli Institute for Theoretical Physics, University
of California, Santa Barbara, California  93106-4030}

\begin{abstract}
Entanglement witnesses (EWs) constitute one  of the most important
entanglement detectors in quantum systems. Nevertheless, their
complete characterization, in particular with respect to the
notion of optimality, is still missing, even in the decomposable
case. Here we show that for any qubit-qunit decomposable EW (DEW)
$W$ the three statements are equivalent: (i) the set of product
vectors obeying $\bra{e,f}W\ket{e,f}=0$ spans the corresponding
Hilbert space, (ii) $W$ is optimal, (iii) $W=Q^{\Gamma}$ with $Q$
denoting a positive operator supported on a completely entangled
subspace (CES) and $\Gamma$ standing for the partial
transposition. While, implications $(i)\Rightarrow(ii)$ and
$(ii)\Rightarrow(iii)$ are known, here we prove that $(iii)$
implies $(i)$. This is a consequence of a more general fact saying
that product vectors orthogonal to any CES in
$\mathbbm{C}^{2}\ot\mathbbm{C}^{n}$ span after partial conjugation
the whole space. On the other hand, already in the case of
$\mathbbm{C}^{3}\ot\mathbbm{C}^{3}$ Hilbert space, there exist
DEWs for which (iii) does not imply (i). Consequently, either (i)
does not imply (ii), or (ii) does not imply (iii), and the above
transparent characterization obeyed by qubit-qunit DEWs, does not
hold in general.
\end{abstract}

\maketitle

\section{Introduction}

Entanglement witnesses (EW) \cite{Horodecki96PLA,Terhal00PLA}
provide one of the best known methods of entanglement detection in
composite (bipartite and multipartite) quantum systems (see the
recent review \cite{GuhneTothReview} for other methods). These are
Hermitian operators which, on one hand, have nonnegative mean
values in all separable states and, on the other hand, they must
have negative mean values in some entangled states.

The particular importance of EWs in detection of entanglement
stems from several facts. First of all, we know that they give
rise to a necessary and sufficient condition for separability
\cite{Horodecki96PLA} (see also \cite{Horodecki01PLA} for the
multipartite case). Precisely, given $\varrho$ is separable if and
only if $\langle W\rangle_{\varrho}\equiv \Tr(W\varrho)\geq 0$ for
all EWs or, equivalently, $\varrho$ is entangled if and only if
$\langle W\rangle_{\varrho}<0$ for at least one such $W$. For the
above reasons it is not feasible to check the "if" part of this
criterion, nevertheless it still gives a strong necessary
condition for separability. Then, as it was first stressed in Ref.
\cite{Terhal00PLA}, since EWs are Hermitian operators, it is clear
that they correspond to some quantum observables and therefore the
above criterion is applicable in experiment (see e.g.
\cite{Barbieri03PRL,Bourennane04PRL}). Finally, there is a whole
bunch of works indicating their quantitative meaning (see e.g.
\cite{BrandaoVianna06IJQI,Brandao05PRA,Audenaert07NJP,Reimpell07PRL,Reimpell08PRA}).
More precisely, mean values of entanglement witnesses not only
serve as entanglement detectors, but also can tell us how much
entangled the state is.

Although the above extensive literature as well as
Refs. \cite{LewensteinKraus00PRA,LewensteinKraus01PRA,Acin01PRL,TothGuhne05PRL,Toth05PRA,
Sarbicki08JPA,Korbicz08PRA,SperlingVogel09PRA,Chruscinski09CMP,Sarbicki09,Skowronek09JMP})
is aiming at studying their properties and providing methods
of construction it seems that still much can and should to be said about EWs. In
particular, complete characterization and classification of EWs is
far from satisfactory. The structure of the so-called optimal EWs
(in the sense of Ref. \cite{LewensteinKraus00PRA}, see also below
for the definition), even in the decomposable case, is still
unknown. The importance of this problem stems from the fact that
the above separability criterion can be restated using optimal EWs
only. This is because every EW which is not optimal can be
optimized \cite{LewensteinKraus00PRA}. Therefore it is of great
importance to characterize the set of EWs with respect to their
optimality. The early attempts to achieve this goal were discussed
already in \cite{LewensteinKraus00PRA}.

The main purpose of this note is to go towards solving the above
problems. We investigate few notions connected to the optimality
of the decomposable entanglement witnesses. In particular, we show
that in the case of qubit-qunit Hilbert spaces, a more exhausting
characterization with respect to optimality can be given.
Precisely, for all qubit-qunit decomposable entanglement witnesses
the three statements are equivalent: (i) $W$ is optimal, (ii)
$W=Q^{\Gamma}$ with $Q$ being a positive operator supported on a
completely entangled subspace (CES) and $\Gamma$ denoting the
partial transposition map\footnote{Note that we do not specify the
subsystem on which the transposition map is applied since our
results are independent of this choice. However, for convenience,
in all the proofs the transposition map is applied to the
lower-dimensional subsystem.}, (iii) the Hilbert space
$\mathbbm{C}^{2}\ot\mathbbm{C}^{n}$ is spanned by product vectors
obeying $\bra{e,f}W\ket{e,f}=0$. We achieve this goal by showing
that product vectors orthogonal to any CES of
$\mathbbm{C}^{2}\ot\mathbbm{C}^{n}$ after partial conjugation
(PC)\footnote{By the partial conjugation of a product vector
$\ket{e,f}$ we mean the complex conjugation of either $\ket{e}$ or
$\ket{f}$. Since our result do not depend on the choice of the
subsystem subject to PC, we do not state explicitly on which
subsystem it acts. Nevertheless, for convenience, in all the
proofs, it is applied to the lower-dimensional subsystem.} span
$\mathbbm{C}^{2}\ot\mathbbm{C}^{n}$. This means that (ii) implies
(iii) and together with already proven facts that (iii) implies
(i) and (i) implies (ii) \cite{LewensteinKraus00PRA}, gives the
above equivalence. The above fact also solves, at least in this
particular case, the long-standing question whether (i) implies
(iii).

Then we study DEWs acting on higher-dimensional Hilbert spaces and
show that already in the simplest case of $3\ot 3$ the above
equivalence appears to be false. Specifically, depending on the
rank of $Q$, (ii) does not always imply (iii). This in turn
implies that either not all witnesses admitting the form
$W=Q^{\Gamma}$ are optimal ((ii) does not imply (i)), or not all
OEWs have the property that product vectors satisfying
$\bra{e,f}W\ket{e,f}=0$ span the corresponding Hilbert space ((i)
does not imply (iii)).

It should be noticed that in the case of indecomposable EWs (IEW),
examples of witnesses for which (i) does not imply (iii) are
already known. A particular example of such a witness comes from
the Choi map \cite{Choi75LAA2}. The latter is extremal in the
convex set of positive maps \cite{ChoiLam} and therefore gives a
optimal EW (see e.g. Ref. \cite{Skowronek09JMP}). On the other
hand, product vectors from $\mathbbm{C}^{3}\ot\mathbbm{C}^{3}$ at
which the witness has zero mean value span a seven dimensional
subspace in $\mathbbm{C}^{3}\ot\mathbbm{C}^{3}$ (see Refs.
\cite{Korbicz08PRA,Sarbicki09}). Recently, using the theory of
convex cones, the geometrical properties of such witnesses have
been studied in Ref. \cite{Sarbicki09}.

The paper is organized as follows. In Sec. \ref{Preliminaries} we
recall all the necessary notions and present, in a concise way,
all we need about optimality of DEW. Then, in Sec. \ref{Results},
we present our main results. We conclude in Sec. \ref{Conclusion}.

\section{Preliminaries}
\label{Preliminaries}

For further benefits let us now recall some definitions and facts
regarding decomposable entanglement witnesses. We give the
definitions of separable states, entanglement witnesses, optimal
and decomposable entanglement witnesses. Then, we shortly remind
what is known regarding relations between optimality and
decomposability of EWs.

In what follows we will be concerned with finite--dimensional
product Hilbert spaces $\mathbbm{C}^{m}\ot\mathbbm{C}^{n}$,
henceforward denoted shortly by $\mathcal{H}_{m,n}$. By
$\mathcal{D}_{m,n}$ and $\mathcal{D}_{m,n}^{\mathrm{sep}}$ we
denote, respectively, the set of all density matrices and
separable density matrices acting on $\mathcal{H}_{m,n}$. In the
case of equal local dimensions $m=n$, we use a single subscript
$m$. Finally, $M_m(\mathbbm{C})$ will denote the set of $m\times
m$ matrices with complex entries.

Following Ref. \cite{Werner89PRA}, we call a density matrix
$\varrho$ acting on $\mathcal{H}_{m,n}$ {\it separable} if it can
be written as
\begin{equation}
\varrho=\sum_{i}p_{i}\proj{a_{i}}\ot\proj{b_{i}},\qquad p_i\geq 0,
\qquad \sum_{i}p_{i}=1,
\end{equation}
where $\ket{a_{i}}$ and $\ket{b_{i}}$ denote some pure states from
$\mathbbm{C}^{m}$ and $\mathbbm{C}^{n}$, respectively.

In 1996, basing on the Hahn-Banach separation theorem (cf.
\cite{book}), an important fact regarding the separability problem
was proven \cite{Horodecki96PLA}. Namely, a state $\varrho$ acting
on $\mathcal{H}_{m,n}$ is entangled if and only if there exists a
Hermitian operator $W\in M_{m}(\mathbbm{C})\ot M_{n}(\mathbbm{C})$
such that $\Tr(W\varrho)<0$ and at the same time $\Tr(W\sigma)\geq
0$ for all $\sigma\in\mathcal{D}_{m,n}^{\mathrm{sep}}$. This fact
gives rise to the following definition. Any Hermitian operator $W$
acting on $\mathcal{H}_{m,n}$ is called {\it entanglement witness}
if it has the properties: (i) its mean value $\langle
W\rangle_{\sigma}$ in any
$\sigma\in\mathcal{D}_{m,n}^{\mathrm{sep}}$ is nonnegative, (ii)
there exists an entangled state $\sigma$ such that $\langle
W\rangle_{\sigma}<0$. Notice, that both the conditions can be
rephrased as follows: (i) $\langle e,f|W|e,f\rangle\geq 0$ for any
pair of vectors $\ket{e}\in\mathbbm{C}^{m}$ and
$\ket{f}\in\mathbbm{C}^{n}$, (ii) $W$ has at least one negative
eigenvalue.

Now, via the Choi-Jamio\l{}kowski isomorphism
\cite{Choi75LAA,Jamiolkowski72RMP}, the theory of positive maps
induces the following partition of EWs
\cite{Woronowicz76RMP,LewensteinKraus00PRA}. An entanglement
witness $W$ is called {\it decomposable} (DEW) if it can be
written as $W=aP+(1-a)Q^{\Gamma}$ with $P,Q\geq 0$ and
$a\in[0,1]$. EWs that do not admit this form are called {\it
indecomposable}.
Notice that the decomposable witnesses detect only states which
have nonpositive partial transposition (NPT). For detection of
entangled states with positive partial transposition we need to
use indecomposable entanglement witnesses.

Let us now pass to the notion of optimality. To this aim we
introduce
\begin{equation}
D_{W}=\{\varrho\in\mathcal{D}_{m,n}|\langle
W\rangle_{\varrho}<0\},
\end{equation}
that is, the set of all entangled states detected by $W$.
Following Ref. \cite{LewensteinKraus00PRA}, we say that given two
EWs $W_i$ $(i=1,2)$, $W_{1}$ is {\it finer} than $W_{2}$ if
$D_{W_{2}}\subseteq D_{W_{1}}$. Then, we say that $W$ is {\it
optimal} if there does not exist any other entanglement witness
which is finer than $W$.

It was shown in Ref. \cite{LewensteinKraus00PRA} that $W_1$ is
finer than $W_2$ if and only if there exists a positive number
$\epsilon$ and a positive operator $P$ such that $W_{1}$ can be
expressed as $W_1=(1-\epsilon)W_{2}+\epsilon P$. This immediately
implies that $W$ is optimal iff for any $\epsilon>0$ and $P\geq
0$, the operator $\widetilde{W}=(1+\epsilon)W-\epsilon P$ is not
an EW. The only candidates for positive operators that can be
subtracted from $W$ according to the above recipe must obey
$PP_W=0$, with
\begin{equation}
P_W=\{\ket{e,f}\in\mathcal{H}_{m,n}|\langle e,f|W|e,f\rangle=0\}.
\end{equation}
This implies a sufficient criterion for optimality of EWs.
Namely, if the set of product vectors $P_W$ spans the
Hilbert space $\mathcal{H}_{m,n}$, the witness $W$ is optimal.
Eventually, application of the above facts to the general form of
DEW allows us to conclude that if a decomposable EW is optimal, it
has to be of the form
\begin{equation}\label{decwitness}
W=Q^{\Gamma},\qquad Q\geq 0,
\end{equation}
where $\supp(Q)$ does not contain any product vectors, or, in
other words, $\supp(Q)$ is a completely entangled subspace (CES)
in $\mathcal{H}_{m,n}$.


\section{Optimality and product vectors in subspaces orthogonal to
completely entangled subspaces} \label{Results}

From the preceding section we know that regarding optimality of
decomposable EWs, two facts hold: (i) if a DEW $W$ is optimal,
then it has to have the form (\ref{decwitness}) and (ii) if $P_W$
corresponding to $W$ spans $\mathcal{H}_{m,n}$, then $W$ is
optimal. One could then ask if the opposite statements are also
true. Or, in other words, if optimality of $W$ is equivalent to
the form (\ref{decwitness}), or to the fact that $P_W$ spans
$\mathcal{H}_{m,n}$.

First we show that in the case of the Hilbert space
$\mathcal{H}_{2,n}$ the fact that DEW $W$ can be written as in Eq.
(\ref{decwitness}), implies that $P_W$ spans $\mathcal{H}_{2,n}$.
This immediately implies that both the above equivalences hold. On
the other hand, we show that already in the $3\ot 3$ case there
are witnesses admitting the form (\ref{decwitness}), but the $P_W$
does not span $\mathcal{H}_{3}$. Consequently, one of the above
equivalences cannot hold. Either not all DEWs of the form
(\ref{decwitness}) are optimal, or optimality does not imply that
$P_W$ spans $\mathcal{H}_{3}$.

Before we start with our proofs, let us notice that since we deal
only with witnesses that admit the form (\ref{decwitness}), the
question about properties of $P_W$ can be seen as the question
about properties of product vectors orthogonal to completely
entangled subspaces. This is a consequence of a simple property of
the transposition map saying that its dual map is again the
transposition map, which allows to conclude that
$\bra{e,f}Q^{\Gamma}\ket{e,f}=\bra{e^{*},f}Q\ket{e^{*},f}$ any
product vector $\ket{e,f}\in\mathcal{H}_{m,n}$. This together with
positivity of $Q$ allows to conclude that $\ket{e,f}$ belongs to
$P_W$ iff $\ket{e^{*},f}\in\mathrm{ker}(Q)$. Thus, in what follows
we can ask a bit more general question, namely if partially
conjugated product vectors orthogonal to a given CES, span the
corresponding Hilbert space. For instance, we will show that for
any CES $V$ of $\mathbbm{C}^{2}\ot\mathbbm{C}^{n}$, the product
vectors belonging to $V^{\perp}$ span $V^{\perp}$, while their
partial conjugations span $V$. Notice that completely entangled
subspaces were recently investigated e.g. in
\cite{Wallach02ContMath,Parthasarathy04,CubittJMP08,WalgateJPA}.
In particular, it was shown that the maximal dimension of CES in
$\mathcal{H}_{m,n}$ is $(m-1)(n-1)$. This translates to the upper
bound on the rank of $Q$ in (\ref{decwitness}), i.e., $r(Q)\leq
(m-1)(n-1)$.

We still need to introduce some more terminology. We say that a
positive operator $Q$ is supported on $\mathcal{H}_{m,n}$ if
$Q_{A}$ and $Q_{B}$\footnote{By $Q_A$ and $Q_B$ we denote
$\Tr_{B}(Q)$ and $\Tr_{A}(Q)$, respectively. Notice that both are
positive.} have ranks $m$ and $n$. Otherwise, if either $Q_{A}$ or
$Q_{B}$ contains some vectors in its kernel, the operator $Q$ acts
effectively on a Hilbert space with smaller dimension. This can be
translated to subspaces of $\mathcal{H}_{m,n}$. We can say that a
given $V$ is supported in $\mathcal{H}_{m,n}$ if the latter
is the "smallest" Hilbert space of which $V$ can be a subspace. In
other words, projector onto this subspace is supported on
$\mathcal{H}_{m,n}$.

Eventually, let $V$ be a subspace of some Hilbert space
$\mathcal{H}$. By $V^{\perp}$ we will be denoting the subspace of
$\mathcal{H}$ of all vectors orthogonal to $V$ (complement of $V$
in $\mathcal{H}$). Also, the notation
\begin{equation}
\mathbbm{C}^{n}\ni\ket{f}=\sum_{i}f_{i}\ket{i}\equiv(f_{0},f_{1},\ldots,f_{n-1})
\end{equation}
will be frequently used.

\subsection{Decomposable witnesses acting on $\mathcal{H}_{2,n}$}

Let us first concentrate on the simplest case of $m=2$.
It follows from the previous discussion that
the maximal dimension of a completely entangled subspace of
$\mathbbm{C}^{2}\ot \mathbbm{C}^{n}$ is $n-1$. For pedagogical purposes
let us start our considerations with this case. Then, we will pass to the cases
of the remaining possible dimensions.\\

\noindent{\bf Lemma 1.} {\it Let $V$ be a $(n-1)$-dimensional CES
of $\mathbbm{C}^{2}\ot\mathbbm{C}^{n}$.
Then there exists a nonsingular $n\times n$ matrix
$A$ such that the family of product vectors
\begin{equation}
\ket{e(\alpha),f(\alpha)}\equiv(1,\alpha)\ot A(1,\alpha,\ldots,\alpha^{n-1}) \qquad
(\alpha\in\mathbbm{C})
\end{equation}
belong to $V^{\perp}$. Moreover, the vectors
$\ket{e(\alpha),f(\alpha)}$ $(\alpha\in\mathbbm{C})$ span
$V^{\perp}$, while $\ket{e^{*}(\alpha),f(\alpha)}$ span
$\mathbbm{C}^{2}\ot\mathbbm{C}^{n}$.\\}

\noindent{\bf Proof.} Let $\ket{\Psi_{i}}$ $(i=1,\ldots,n-1)$
denote the linearly independent vectors spanning $V$. All of them
have to be entangled as otherwise there would exist a product
vector in $V$. This means that they can be expressed as
\begin{equation}\label{PsiForm}
\ket{\Psi_i}=\ket{0}\ket{\psi_{0}^{(i)}}+\ket{1}\ket{\psi_{1}^{(i)}}\qquad
(i=1,\ldots,k),
\end{equation}
with nonzero vectors $\ket{\psi_{j}^{(i)}}$ $(i=1,\ldots,k;j=0,1)$
such that the vectors in each pair $\ket{\psi_{j}^{(i)}}$
$(j=0,1)$ are linearly independent. Also, it is easy to see that
the vectors in both the sets, $\{\ket{\psi_{0}^{(i)}}\}_{i}^{n-1}$
and $\{\ket{\psi_{1}^{(i)}}\}_{i}^{n-1}$, are linearly
independent. Otherwise in both cases it is possible to find a
product vector in $V$.

Let us now look for the product vectors $\ket{e,f}$ orthogonal to
$V$, where we take $\ket{e}=(1,\alpha)\in\mathbbm{C}^{2}$ with
$\alpha\in\mathbbm{C}$ and arbitrary
$\ket{f}=(f_0,\ldots,f_{n-1})\in\mathbbm{C}^{n}$. The
orthogonality conditions to $\ket{\Psi_i}$ $(i=1,\ldots,n-1)$ give
us the set of $n-1$ linear homogenous equations
\begin{equation}
\langle\Psi_{i}|e,f\rangle=0\qquad (i=1,\ldots,n-1).
\end{equation}
for $n$ variables $f_i$. In order to solve it, we can fix one of
the variables, say $f_0=1$, getting a system of $n-1$
inhomogenous equations for $n-1$ variables.
It can easily be solved and the solution is given by
\begin{equation}
f_{i}(\alpha)=\frac{R_{i}(\alpha)}{R(\alpha)}\qquad
(i=1,\ldots,n-1),\label{malutka}
\end{equation}
where $R_{i}$ and $R$ are polynomials in $\alpha$ of degree at
most $n-1$. Moreover, since the vectors $\ket{\psi_{1}^{(i)}}$
$(i=1,\ldots,n)$ are linearly independent, the degree of the
polynomial $R$ is exactly $n-1$. Consequently, the product vectors
in $V^{\perp}$ we look for, take the generic form
\begin{equation}\label{product_vectors}
\ket{e(\alpha),f(\alpha)}=(1,\alpha)\ot(R(\alpha),R_{1}(\alpha),\ldots,R_{n-1}(\alpha))
\quad (\alpha\in\mathbbm{C}).
\end{equation}
Note that we have multiplied above everything by $R(\alpha)$, so
that the expression (\ref{product_vectors}) is valid also for
$R(\alpha)=0$, while the expression (\ref{malutka}) only when
$R(\alpha)\ne 0$. Nevertheless, by continuity or by local change
of the basis one shows that the vectors (\ref{product_vectors})
for $\alpha$ being the roots of  $R$ are also orthogonal to
$\ket{\Psi_i}$.

For further purposes, let us denote by
$V_{\mathrm{sep}}^{\perp}$ the subspace of $V^{\perp}$ spanned by
all the vectors (\ref{product_vectors}).

The assumption that $V$ does not contain any product vector
implies that all the polynomials $R$, $R_{i}$ are linearly
independent. In order to see it explicitly, let us assume that
only $k<n$ of them are linearly independent. Then, there has to
exist $n-k$ vectors $\ket{\xi_i}$ $(i=1,\ldots,n-k)$ that are
orthogonal to the subspace of $\mathbbm{C}^{n}$ spanned by
$\ket{f(\alpha)}$ $(\alpha\in\mathbbm{C})$. Moreover, for any
$\ket{h}\in\mathbbm{C}^{2}$, vectors $\ket{h}\ket{\xi_i}$
$(i=1,\ldots,n-k)$ are orthogonal to $V^{\perp}_{\mathrm{sep}}$.
In what follows we show that among the latter there exists at
least one product vector which is orthogonal to $V^{\perp}$ and
thus has to be in $V$ leading to the contradiction with the
assumption that $V$ is a CES.

For this purpose, let us notice that vectors
(\ref{product_vectors}) span $(k+1)$-dimensional subspace in
$V^{\perp}$. As a result, there exists a set of $n-k$ vectors
$\ket{\omega_i}\in V^{\perp}$ $(i=1,\ldots,n-k)$, which are
orthogonal to all $\ket{e(\alpha),f(\alpha)}$. Now, we take the
following product vector
\begin{equation}
\ket{\eta}=(\ket{0}+\gamma\ket{1})\ot\sum_{i=1}^{n-k}b_{i}\ket{\xi_{i}}
\end{equation}
with $\gamma\in\mathbbm{C}$ and $b_i\in\mathbbm{C}$ being some
parameters to be determined. Obviously, $\ket{\eta}$ is already
orthogonal to $V_{\mathrm{sep}}^{\perp}$. Orthogonality conditions
$\langle\omega_i|\eta\rangle=0$ $(i=1,\ldots,n-k)$ give us the
system of $n-k$ homogenous equations for $n-k$ variables $b_i$ of
the form $(M_{1}+\gamma M_{2})\ket{b}=0$ with
$\ket{b}=(b_{1},\ldots,b_{n-k})$ and $M_{i}$ being some matrices.
It has a nontrivial solution only if $\det(M_{1}+\gamma M_{2})$
vanishes. The latter is a polynomial in $\gamma$ of at most
$(n-k)$th degree and obviously the corresponding equation is
soluble in the complex field. Consequently, we have product
vectors belonging to $V$, which is in a contradiction with the
assumption that $V$ is a CES. Thus, $R$, $R_i$ $(i=1,\ldots,n-1)$
are linearly independent. This in turn means that there exists a
nonsingular transformation $A:\mathbbm{C}^{n}\to\mathbbm{C}^{n}$
such that $\ket{f(\alpha)}=A(1,\alpha,\ldots,\alpha^{n-1})$ for
any $\alpha\in\mathbbm{C}$.

On the other hand, it is easy to see that vectors
\begin{equation}\label{vectors}
(1,\alpha)\ot(1,\alpha,\ldots,\alpha^{n-1})\qquad (\alpha\in\mathbbm{C})
\end{equation}
span $(n+1)$-dimensional subspace of
$\mathbbm{C}^{2}\ot\mathbbm{C}^{n}$, while their PCs, that is,
\begin{equation}\label{vectors2}
(1,\alpha^{*})\ot(1,\alpha,\ldots,\alpha^{n-1})\qquad (\alpha\in\mathbbm{C}),
\end{equation}
span the whole $\mathbbm{C}^{2}\ot\mathbbm{C}^{n}$. In the first
case this is because among $2n$ monomials in $\alpha$ appearing in
Eq. (\ref{vectors}), $n+1$ are linearly independent. In the second
case, $\alpha^{*}$ is linearly independent of any polynomial in
$\alpha$ and thus we have $2n$ linearly independent polynomials in
(\ref{vectors2}). Therefore, since $A$ is of full rank, vectors
$\ket{e(\alpha)}\ot A\ket{f(\alpha)}$ $(\alpha\in\mathbbm{C})$
span $V$, while $\ket{e^{*}(\alpha)}\ot A\ket{f(\alpha)}$ the
whole $\mathcal{H}_{2,n}$. This finishes the proof. $\blacksquare$

Let us now pass to the remaining cases with respect to the
dimension of $V$.\\

\noindent{\bf Lemma 2.} {\it Let $V$ be a $k<n-1$--dimensional CES
of $\mathbbm{C}^{2}\ot\mathbbm{C}^{n}$.
Then there exists a nonsingular transformation
$A$, such that the vectors
\begin{eqnarray}\label{lemma2}
\fl(1,\alpha)\ot
A(R(\alpha,\beta),\beta_{1}R(\alpha,\beta),\ldots,\beta_{n-k-1}R(\alpha,\beta),
R_{1}(\alpha,\beta),\ldots,R_{k}(\alpha,\beta))\nonumber\\
\hspace{6cm}(\alpha,\beta_1,\ldots,\beta_{n-k-1}\in\mathbbm{C})
\end{eqnarray}
span $V^{\perp}$, while their PCs span
$\mathbbm{C}^{2}\ot\mathbbm{C}^{n}$. Here $\beta\equiv
(\beta_1,\ldots,\beta_{n-k-1})$, $R(\alpha,\beta)$ and
$R_{i}(\alpha,\beta)$ are polynomials of at most $k$th degree in
$\alpha$ and first degree in $\beta_i$ $(i=1,\ldots,n-k-1)$.\\}

\noindent{\bf Proof.} We can follow the same reasoning as in the
proof of Lemma 1. Now, we have $k$ entangled vectors
$\ket{\Psi_i}$ spanning $V$ which can be written as in Eq.
(\ref{PsiForm}). For the same reason as before both sets
$\{\ket{\psi_{0}^{(i)}}\}_{i=0}^{k-1}$ and
$\{\ket{\psi_{1}^{(i)}}\}_{i=0}^{k-1}$ are linearly independent.
Therefore we can always find a nonsingular transformation
$\widetilde{A}:\mathbbm{C}^{n}\to\mathbbm{C}^{n}$ such that
$\widetilde{A}\ket{\psi_{1}^{(i)}}=\ket{n-k-1+i}$
$(i=1,\ldots,k)$.

Let us now consider the locally transformed subspace
$\widetilde{V}=(\mathbbm{1}_2\ot \widetilde{A})V(\mathbbm{1}_2\ot \widetilde{A}^{\dagger})$,
which is also a CES, and look for the separable vectors
belonging to $\widetilde{V}^{\perp}$ and taking the following form
\begin{equation}
(1,\alpha)\ot(1,\beta_{1},\ldots,\beta_{n-k-1},f_{1},\ldots,f_{k}),
\end{equation}
where $\beta_{i}\in\mathbbm{C}$ are free parameters
and $f_{i}$ $(i=1,\ldots,k)$ are to be determined. Orthogonality
conditions to $k$ vectors spanning $\widetilde{V}$, i.e.,
$\ket{\widetilde{\Psi}_i}=\mathbbm{1}_{2}\ot\widetilde{A}\ket{\Psi_i}$,
lead us to the following inhomogenous linear equations
\begin{equation}
\sum_{j=1}^{k}f_{j}\left(\langle\widetilde{\psi}_{0}^{(i)}|n-k-1+j\rangle
+\alpha\delta_{ij}\right)
=x_{i}(\alpha,\beta)\quad\; (i=1,\ldots,k),
\end{equation}
where $x_{i}(\alpha,\beta)$ are polynomials of the first degree in $\alpha$
and all $\beta$s.

Following the same reasoning as in the proof of lemma 1, one obtains
the product vectors orthogonal to $\ket{\widetilde{\Psi}_i}$ in the form
\begin{equation}\label{separableVec3}
\fl \ket{e(\alpha),f(\alpha,\beta)}=(1,\alpha)\ot
(R,\beta_1R,\ldots,\beta_{n-k-1}R, R_{1},\ldots,R_{k})\qquad
(\alpha,\beta_i\in\mathbbm{C}),
\end{equation}
where $R_i$ and $R$ are polynomials of degree at most $k$ in
$\alpha$ and one in $\beta$s (for brevity we omitted arguments of
$R$ and $R_i$ in (\ref{separableVec3})). Moreover, due to the
already mentioned fact that the vectors $\ket{\psi_{1}^{(i)}}$
$(i=0,\ldots,k-1)$ are linearly independent, the highest power of
$\alpha$ in $R$ is exactly $k$.

Let us now show that the polynomials $R$, $\beta_{i}R$
$(i=1,\ldots,n-k-1)$, and $R_{i}$ $(i=1,\ldots,k)$ are linearly
independent. For this purpose, let us assume that only $m<n$ of
them are linearly independent. It is clear that $m\geq n-k$ as the
monomials $1$ and $\beta_i$ $(i=1,\ldots,n-k-1)$ are by the very
definition linearly independent and therefore we can denote
$m=n-k+l$ with $l=1,\ldots,k$. Consequently, there exist $k-l$
vectors $\ket{\widetilde{\xi}_i}\in\mathbbm{C}^{n}$ orthogonal to
the subspace spanned by $\ket{f(\alpha,\beta)}$
$(\alpha,\beta_i\in\mathbbm{C})$.

On the other hand, since $R$ is of $k$th degree in $\alpha$ and
the above $m$ polynomials are of degree at most $k$ in $\alpha$,
they, together with $n-k$ polynomials $\alpha R(\alpha,\beta)$ and
$\alpha\beta_{i}R(\alpha,\beta)$ $(i=1,\ldots,n-k-1)$, constitute
the set of $2(n-k)+l$ linearly independent polynomials.
This implies that the vectors (\ref{separableVec3}) 
span at least $2(n-k)+l$-dimensional subspace in
$\widetilde{V}^{\perp}$. In the worst case scenario, i.e., when
this dimension is exactly $2(n-k)+l$ we have $k-l$ linearly
independent $\ket{\widetilde{\omega}_i}\in \widetilde{V}^{\perp}$
which are orthogonal to all vectors (\ref{separableVec3}). Then,
following the same reasoning as in the proof of lemma 1, we can
show that there are product vectors in $\widetilde{V}$, which
contradicts the fact that $\widetilde{V}$ is CES.

In conclusion, all the polynomials $R$, $\beta_i R$
$(i=1,\ldots,n-k-1)$ and $R_i$ $(i=1,\ldots,k)$ are linearly
independent. As a result, these $n$ polynomials together with
$n-k$ polynomials $\alpha R$ and $\alpha\beta_i R$
$(i=1,\ldots,n-k-1)$ constitute the set of $2n-k$ linearly
polynomials and therefore the continuous set of product vectors
$\ket{e(\alpha),f(\alpha,\beta_i)}$ in Eq. (\ref{separableVec3})
span $\widetilde{V}$. Also, for the same reason as before, the
partially conjugated vectors
\begin{equation}
(1,\alpha^{*})\ot (R,\beta_1R,\ldots,\beta_{n-k-1}
R,R_1,\ldots,R_k)
\end{equation}
span $\mathcal{H}_{2,n}$.

Eventually, putting $A=(\widetilde{A}^{-1})^{\dagger}$, we see
that the vectors (\ref{lemma2}) span $V^{\perp}$, while their PCs
span $\mathcal{H}_{2,n}$. This completes the proof. $\blacksquare$\\

The above lemmas together with the previously known results
allow us to prove the following theorem.\\

\noindent{\bf Theorem 1.} {\it Let $W$ be a decomposable witness
acting on $\mathcal{H}_{2,n}$. The the following statements are
equivalent:}
\begin{description}
{\it   \item[(i)] $W=Q^{\Gamma}$, where $Q\geq 0$ and $\supp(Q)$
is a CES in $\mathcal{H}_{2,n}$,
  \item[(ii)] $P_W$ spans $\mathbbm{C}^{2}\ot\mathbbm{C}^{n}$,
  \item[(iii)] $W$ is optimal.}
\end{description}
\noindent{\bf Proof.} The implications $(ii)\Rightarrow (iii)$ and $(iii)\Rightarrow(i)$
were proven in Ref. \cite{LewensteinKraus00PRA}. The implication $(i)\Rightarrow (ii)$
follows from the above lemmas. $\blacksquare$\\

Let us illustrate the above discussion with a simple example. Let
us consider a witness $W=Q^{\Gamma}$ with $Q$ supported on a
$(n-1)$--dimensional subspace $V$ of
$\mathbbm{C}^{2}\ot\mathbbm{C}^{n}$ spanned by the following
vectors
\begin{equation}
\ket{\Psi_i}=(1/\sqrt{2})(\ket{0,i}-\ket{1,i-1}) \qquad (i=1,\ldots,n-1)
\end{equation}
The subspace $V$ does not contain any product vector because, as
one can directly check, there does not exist product vector
orthogonal to
$V^{\perp}=\mathrm{span}\{\ket{00},\ket{1,n-1},(1/\sqrt{2})(\ket{0,i}+\ket{1,i-1})\;(i=1,\ldots,n-1)\}$.
Then the separable vectors spanning $V^{\perp}$ are given by
(\ref{vectors}) and, as already mentioned, they span
$\mathbbm{C}^{2}\ot\mathbbm{C}^{n}$

\subsection{Decomposable witnesses acting on $\mathcal{H}_{3}$}

Here we show that the simple characterization we proved in theorem
1 for $2\ot n$ decomposable witnesses does not hold for some of
witnesses acting already on $\mathbbm{C}^{3}\ot\mathbbm{C}^{3}$.
Precisely, we will see that for witnesses (\ref{decwitness}) with
$r(Q)=1,2$, the analog of the above theorem also holds, while
there are witnesses with $r(Q)=3,4$ such that the separable
vectors from the corresponding $P_W$s do not span $\mathcal{H}_3$.

Let us start from the case of $r(Q)=1$. Here we have a bit more general
fact (see also Ref. \cite{Skowronek09JMP} for a proof of optimality
via extremality). Then, we will consider the case of $r(Q)=2$.\\

\noindent{\bf Lemma 3.} {\it Let $W=\proj{\psi}^{\Gamma}$, where
$\ket{\psi}$ is an entangled pure state from $\mathcal{H}_m$.
Then, the statements (i), (ii), and (iii) from theorem 1 (accordingly reformulated)
are equivalent.\\}

\noindent{\bf Proof.} As previously, implications $(ii)\Rightarrow (iii)$ and $(iii)\Rightarrow(i)$
follow from Ref. \cite{LewensteinKraus00PRA}. Below we prove that $(i)$ implies
$(ii)$.

The Schmidt decomposition of $\ket{\psi}$ reads
\begin{equation}
\ket{\psi}=\sum_{i=0}^{s-1}\sqrt{\mu_{i}}\ket{ii},
\end{equation}
where $\mu_i\geq0$ and $s\leq m$ denotes the Schmidt rank of
$\ket{\psi}$. Without any loss of generality we can assume that
$s=m$. Then, by a local full rank transformation we can bring
$\ket{\psi}$ to the maximally entangled state
$\ket{\psi_{m}^{+}}$. The product vectors orthogonal to the latter
are of the form $\ket{e}\ket{f}$, where $\ket{e}$ is arbitrary
vector from $\mathbbm{C}^{m}$ and $\ket{f}\in\mathbbm{C}^{m}$ is
any vector orthogonal to $\ket{e}$. Then, this class of vectors
after PC span $\mathbbm{C}^{m}\ot\mathbbm{C}^{m}$ (cf. Ref.
\cite{Breuer06PRL}). $\blacksquare$

Note, that with a bit more effort the above lemma can be
generalized to any witness $W=\proj{\psi}^{\Gamma}$ acting on
$\mathbbm{C}^{m}\ot\mathbbm{C}^{n}$.\\

\noindent{\bf Lemma 4.} {\it Let $V$ be a CES of
$\mathcal{H}_{3}$ with $\dim V=2$. Then the product
vectors from $V^{\perp}$, when partially conjugated, span
$\mathcal{H}_{3}$.\\}

\noindent{\bf Proof.} Let $\ket{\Psi_i}$ $(i=1,2)$ be two linearly
independent vectors spanning $V$. Clearly, we can assume that at
least one of these vectors, say $\ket{\Psi_1}$, is of Schmidt rank
two. By a local unitary operations it can be brought to
$\ket{\widetilde{\Psi}_1}=\ket{00}+\ket{11}$.

Let us now look for the product vectors orthogonal to $V$ of the
form $(1,\alpha,\beta)\ot (f_0,f_1,f_2)$
$(\alpha,\beta\in\mathbbm{C})$.
From the orthogonality conditions to the transformed vectors
$\ket{\widetilde{\Psi}_i}$ $(i=1,2)$ one infers that they take the
form
\begin{equation}\label{rownanie}
(1,\alpha,\beta)\ot (-\alpha R(\alpha,\beta),R(\alpha,\beta),R_1(\alpha,\beta)),
\end{equation}
with $R$ and $R_1$ being polynomials in $\alpha$ and $\beta$. Let
us now show that the three polynomials $R$, $\alpha R$, and $R_1$
are linearly independent (the first two already are). To this end
we can follow the approach already used in the previous lemmas.
Assume that $R_1$ is linearly dependent on $R$ and $\alpha R$.
Then, there exist a vector $\ket{\xi}\in\mathbbm{C}^3$ orthogonal
to every $(-\alpha
R(\alpha,\beta),R(\alpha,\beta),R_1(\alpha,\beta))$
$(\alpha,\beta\in\mathbbm{C})$ and consequently any vector
$\ket{h}\ket{\xi}$ with arbitrary $\ket{h}\in\mathbbm{C}^3$ is
orthogonal to the vectors (\ref{rownanie}). On the other hand, one
immediately sees that the latter span five-dimensional subspace in
$V^{\perp}$. This means that since $\dim V^{\perp}=7$, there exist
two vectors $\ket{\omega_i}\in V^{\perp}$ $(i=1,2)$ orthogonal to
all vectors in (\ref{rownanie}). It is then clear that among the
two-parameter class $\ket{h}\ket{\xi}$ there exist at least one
vector orthogonal to both $\ket{\omega_i}$ $(i=1,2)$, implying
the existence of a product vector in $V$. This is, however, in a
contradiction with the assumption that $V$ is a CES.

Since then $R$, $\alpha R$, and $\alpha R$ are linearly independent,
the partially conjugated vectors
\begin{equation}
(1,\alpha^{*},\beta^{*})\ot (-\alpha R(\alpha,\beta),R(\alpha,\beta),R_1(\alpha,\beta)) \qquad (\alpha,\beta\in\mathbbm{C})
\end{equation}
certainly span $\mathcal{H}_3$. $\blacksquare$ \\

Basing on the above lemma 3 and lemma 4, we can now formulate the
analog of theorem 1 for some of DEWs acting on $\mathcal{H}_3$.\\

\noindent{\bf Theorem 2.} {\it Let $W$ be a decomposable witness
acting on $\mathcal{H}_{3}$. Then the following conditions are equivalent:}
\begin{description}
  \item[(i)]   $W=Q^{\Gamma}$ with $Q\geq 0$ such that $r(Q)=1,2$ and $\supp(Q)$ being a CES,
  \item[(ii)]  $P_W$ spans $\mathcal{H}_{3}$,
  \item[(iii)] $W$ is optimal.\\
\end{description}

\noindent{\bf Proof.} The implications $(ii)\Rightarrow (iii)$ and $(iii)\Rightarrow(i)$
were proven in Ref. \cite{LewensteinKraus00PRA}. The implication $(i)\Rightarrow (ii)$
follows from lemma 3 and 4. $\blacksquare$\\

Still the cases of $\dim V=3,4$ remain untouched. As we will see
shortly, it is possible to provide examples of three and
four-dimensional CESs supported in $\mathcal{H}_{3}$ such that
their complements, $V^{\perp}$s, contain product vectors, which, when
partially conjugated, do not span $\mathcal{H}_{3}$. While, due to the fact that
five-dimensional subspaces of $\mathcal{H}_3$ have generically
six product vectors (cf. \cite{WalgateJPA,Leinaas2}), the existence
of such three-dimensional CESs for which the product
vectors from their complements do not, when partially transposed,
span $\mathcal{H}_3$ is surprising and interesting. This
implies that there are DEWs (\ref{decwitness}) with $r(Q)=3$ such
that $P_W$s, even if containing continuous classes of product vectors, do not span
$\mathcal{H}_3$. Among such EWs one may look for the analogs of
the aforementioned Choi-like witnesses (optimal witnesses
whose $P_W$s do not span the corresponding Hilbert space)
already known to exist among the indecomposable EWs \cite{Choi75LAA2}.
Still, however, we cannot prove their optimality. On the other hand, it is
possible to provide examples of witnesses (\ref{decwitness}) (thus
also CESs) with $r(Q)=3,4$ such that their $P_W$s do span
$\mathcal{H}_3$.

In the first, three-dimensional case let us consider the subspace
$V_1$ spanned by the following (unnormalized) vectors
\begin{equation}
\ket{01}+\ket{10},\quad
\ket{02}+\ket{20},\quad
\ket{1}(a\ket{1}+b\ket{2})+\ket{2}(a_{2}\ket{1}+b_2\ket{2}).
\end{equation}
For complex $a_i,b_i$ $(i=1,2)$ satisfying $ab_2\neq a_2b$,
$V_1$ does not contain any product vector. Then,
under the conditions that $(a_2+b)^{2}=4ab_2$ and
$b_2\neq 0$, one shows that all the product vectors in $V_1^{\perp}$
are of the form
\begin{equation}
(1,\alpha,\lambda\alpha)\ot (1,-\alpha,-\lambda\alpha)
\end{equation}
and
\begin{equation}
(0,1,\alpha)\ot (0,b+b_2\alpha,-a-a_2\alpha),
\end{equation}
where $\lambda=-(b+a_2)/2b_2$. Direct check allows to conclude
that both classes after the partial conjugation span only a
seven-dimensional subspace in $\mathcal{H}_3$. Finally, since there do not exist
PPT entangled states acting on $\mathcal{H}_3$ of rank three, any positive $Q$
with $r(Q)=3$ and supported on this subspace has to be NPT, thus giving
rise to a proper witness.

In the four-dimensional case the problem of existence of EWs
(\ref{decwitness}) for which $P_W$s do not span the corresponding
Hilbert space is very much related to the results of
\cite{WalgateJPA,Leinaas2}. In particular, five-dimensional
subspaces in $\mathcal{H}_3$ contain generically six product
vectors (five of them are linearly independent), and obviously
cannot span, when partially conjugated, $\mathcal{H}_3$. In order
to provide a particular example of an EW (\ref{decwitness}) with
$r(Q)=4$, one may consider a CES orthogonal to some unextendible
product basis (UPB)\footnote{Following \cite{UPB} we say that a
set of product vectors from some product Hilbert space
$\mathcal{H}$ is unextendible product basis if the vectors are
orthogonal and there does not exist any other product vector in
$\mathcal{H}$ orthogonal to all of them. Skipping the
orthogonality condition we get nonorthogonal UPB.} \cite{UPB} (see
also Ref. \cite{Leinaas}). To this end, let us take one of the
five-elements UPBs from $\mathcal{H}_3$ given in \cite{UPB},
called PYRAMID:
\begin{equation}\label{Pyramid}
\ket{\psi_i}=\ket{\phi_{i}}\ket{\phi_{2i\mathrm{mod}5}}\qquad
(i=0,\ldots,4)
\end{equation}
with
\begin{equation}
\ket{\phi_i}=N\left[\cos\left(\frac{2\pi i}{5}\right)
\ket{0}+\sin\left(\frac{2\pi i}{5}\right)
\ket{1}+h_{+}\ket{2}\right],
\end{equation}
where $h_{\pm}=(1/2)\sqrt{\sqrt{5}\pm 1}$ and
$N=2/\sqrt{5+\sqrt{5}}$. The subspace
orthogonal to these vectors is spanned by orthogonal vectors of
Schmidt rank two given by
\begin{eqnarray}
\eta\ket{10}+(\eta\ket{0}+2h_{-}^{2}\ket{2})\ket{1},\qquad -\eta\ket{10}+\ket{1}(\eta\ket{0}+2h_{-}^{2}\ket{2}),\nonumber\\
(-h_{-}\ket{0}+4h_{-}^{2}\ket{2})\ket{0}-h_{-}\ket{11},\qquad
\ket{0}(h_{-}\ket{0}+4h_{-}^{2}\ket{2})+h_{-}\ket{11},\nonumber\\
\end{eqnarray}
where $\eta=1/2h_{+}$.

Taking convex combination of projectors
onto this vectors, denoted $\mathcal{P}_i$ $(i=1,2,3,4)$,
with equal weights we obviously get PPT entangled state. However, by
appropriate changing these weights we get a positive operator $Q$
which is NPT. For instance we can consider the following
one-parameter family of $Q$s
\begin{equation}
Q(r)=r\left(\mathcal{P}_1+\mathcal{P}_2\right)+(1/2)(1-2r)\left(\mathcal{P}_3+\mathcal{P}_{4}\right)
\quad (0\leq r\leq 1/2).
\end{equation}
It is easy to check that $Q(r)$ is NPT except for $r=1/4$.

In spite of the above examples it is still possible to provide
three and four-dimensional CESs such that the product vectors in
their complements do span, after PC, $\mathcal{H}_3$. Note that
generically for the three-dimensional CESs of $\mathcal{H}_3$ this
is the case. Let us then consider the following subspace:

\begin{equation}
V_2=\spa\{\ket{01}-\ket{10},\ket{02}-\ket{20},\ket{12}-\ket{21},\ket{02}+\ket{20}-\ket{11}\}.
\end{equation}
Note that $V_2$ contains the antisymmetric subspace of
$\mathcal{H}_3$ and the fourth vector spanning it (which is of
Schmidt rank three) belongs to the symmetric subspace of
$\mathcal{H}_3$. It is clear that $V_2$ is supported in
$\mathcal{H}_{3}$ and it does not contain any product vectors. In
order to see it explicitly, assume that some $\ket{e,f}$ can be
written as a linear combination of all these vectors. Application
of the swap operator to $\ket{e,f}$ gives $\ket{f,e}$. On the
other hand, it changes the sign before first three vectors
spanning $V_2$ and therefore one sees that
$\ket{02}+\ket{20}-\ket{11}$ is proportional to
$\ket{e,f}+\ket{f,e}$ which contradicts the fact that it has
Schmidt rank three.

It is now easy to see that the product vectors
\begin{equation}
(1,\alpha,\alpha^{2}/2)\ot(1,\alpha,\alpha^{2}/2)\qquad (\alpha\in\mathbbm{C})
\end{equation}
are orthogonal to $V_2$ and their PC span $\mathcal{H}_{3}$.

\section{Conclusion}
\label{Conclusion}

Let us shortly summarize the obtained results and sketch lines of
further possible research.

Entanglement witnesses give one of the most relevant tools in the
theory of entanglement. Their characterization is therefore of a
great interest. In this note we have focused on the simpler case
of decomposable entanglement witnesses and investigated couple of
issues related to the notion of optimality. In the $2\ot n$ case,
more profound characterization can be given to DEWs. Together with
Ref. \cite{LewensteinKraus00PRA}, our results show that a given
DEW $W$ is optimal iff the corresponding $P_W$ spans
$\mathcal{H}_{2,n}$. Then, the latter holds iff $W=Q^{\Gamma}$
with positive $Q$ supported on some CES. Interestingly, such
transparent characterization does not hold already in the case of
DEWs acting on $\mathcal{H}_{3}$.
Precisely, although for all such DEWs with $r(Q) = 1,2$ the above
equivalences also hold, there exist DEWs with $r(Q) = 3,4$ such
that the product vectors from the corresponding PWs do not span
$\mathcal{H}_3$. This in general means that either not all
witnesses taking the form (\ref{decwitness}) with $Q$ supported on
a CES are optimal, or that optimality of a DEWs $W$ does not
necessarily mean that its $P_W$ spans the corresponding Hilbert
space.

Obviously the obtained results do not complete the
characterization of DEWs, even in the two-qutrit case. In
particular, the complete analysis of the cases when $r(Q)=3,4$ is
missing. Even if for $r(Q) = 3$, generically PWs of DEWs
(\ref{decwitness}) span $\mathcal{H}_3$, it is possible to find
examples of DEWs, as the one provided above, for which this is not
the case. One task would be to characterize such witnesses and
check if some of them are optimal. This would prove that also in
the case of DEWs optimality does not imply that $P_W$ spans the
Hilbert space on which $W$ acts. Let us remind that the existence
of indecomposable EWs having this property is already known
\cite{Choi75LAA2,Korbicz08PRA,Sarbicki09}.

On the other hand, in the case of $r(Q)=4$ is follows from e.g.
\cite{WalgateJPA,Leinaas2} that almost all four-dimensional
subspaces of $\mathcal{H}_3$ have only six product vectors in
their complement meaning that generic $P_W$s of DEW
(\ref{decwitness}) with $r(Q)=4$ do not span $\mathcal{H}_3$.
Nevertheless there exist CESs, as for instance the one presented
above, such that product vectors in their complement span
$\mathcal{H}_3$. Again, it seems interesting to characterize these
CESs.

Then, one could ask the same questions in the case of higher-dimensional
Hilbert spaces $\mathcal{H}_{m,n}$ and, finally, similar analysis is
missing in the case of indecomposable entanglement witnesses.

\section*{Acknowledgments} Discussions with G. Sarbicki and
J. Stasi\'nska are gratefully acknowledged.
This work was supported by EU IP AQUTE, Spanish MINCIN project
FIS2008-00784 (TOQATA), Consolider Ingenio 2010 QOIT, EU STREP
project NAMEQUAM, the Alexander von Humboldt Foundation. R. A.
acknowledges the kind hospitality of the Department of Mathematics
of University of Arizona, where part of this work was done.
M. L. acknowledges NFS Grant PHY005-51164.

\end{document}